\documentclass[aps,prb,twocolumn,superscriptaddress]{revtex4-2}

\usepackage[utf8]{inputenc}
\usepackage{graphicx}
\usepackage{placeins}
\usepackage{amsmath}
\usepackage{amssymb}
\usepackage{braket}
\usepackage{xcolor}
\usepackage{nicefrac}
\usepackage{multirow}
\usepackage[%
  colorlinks=true,
  urlcolor=blue,
  linkcolor=blue,
  citecolor=blue
]{hyperref}
\usepackage{dcolumn}
\usepackage{bm}
\usepackage{epsfig}
\usepackage{latexsym}
\usepackage{amsfonts}
\usepackage{hyperref}
\usepackage{rotating}
\usepackage[normalem]{ulem}

\begin{document}

\title{Bulk and surface electronic structure of Bi$_4$Te$_3$ from $GW$ calculations and photoemission 
experiments }

\author{Dmitrii Nabok}
\affiliation{Peter Gr\"{u}nberg Institut and Institut for Advanced Simulation, Forschungszentrum J\"{u}lich and JARA, 
52425 J\"{u}lich, Germany}

\author{Murat Tas}
\affiliation{Department of Physics, Gebze Technical University, Kocaeli 41400, Turkey}

\author{Shotaro Kusaka}
\affiliation{Department of Physics, Tokyo Institute of Technology, 2-12-1 Ookayama, Meguro-ku, Tokyo 
152-8551, Japan}

\author{Engin Durgun}
\affiliation{UNAM - National Nanotechnology Research Center and Institute of Materials Science and 
Nanotechnology, Bilkent University, Ankara 06800, Turkey}

\author{Christoph Friedrich}
\author{Gustav Bihlmayer}
\author{Stefan Bl\"{u}gel}
\affiliation{Peter Gr\"{u}nberg Institut and Institut for Advanced Simulation, Forschungszentrum J\"{u}lich and JARA, 
52425 J\"{u}lich, Germany}

\author{Toru Hirahara}
\affiliation{Department of Physics, Tokyo Institute of Technology, 2-12-1 Ookayama, Meguro-ku, Tokyo 
152-8551, Japan}

\author{Irene Aguilera}
\email[Corresponding author: ]{i.g.aguilerabonet@uva.nl}
\altaffiliation[Present address: ]{Institute of Physics, University of Amsterdam, 1012WX Amsterdam, Netherlands}
\affiliation{Peter Gr\"{u}nberg Institut and Institut for Advanced Simulation, Forschungszentrum J\"{u}lich and JARA, 
52425 J\"{u}lich, Germany}

\begin{abstract}
We present a combined theoretical and experimental study of the electronic structure of stoichiometric 
Bi$_4$Te$_3$, a natural superlattice of alternating Bi$_2$Te$_3$ quintuple layers and Bi bilayers. In 
contrast to the related semiconducting compounds Bi$_2$Te$_3$ and Bi$_1$Te$_1$, density functional theory 
predicts Bi$_4$Te$_3$ to be a semimetal. In this work, we compute the quasiparticle electronic 
structure of Bi$_4$Te$_3$ in the framework of the $GW$ approximation  
within many-body perturbation theory. 
The quasiparticle corrections are found to modify the dispersion of the valence and conduction bands 
in the vicinity of the Fermi energy, leading to the opening of a small indirect band gap. Based on the 
analysis of the eigenstates, Bi$_4$Te$_3$ is classified as a dual topological insulator with bulk 
topological invariants $\mathbb{Z}_2$ (1;111) and magnetic mirror Chern number $n_M=1$. The bulk $GW$ 
results are used to build a Wannier-functions based tight-binding Hamiltonian that is further applied 
to study the electronic properties of the (111) surface. The comparison with our angle-resolved 
photoemission measurements shows excellent agreement between the computed and measured surface 
states and indicates the 
dual topological nature of Bi$_4$Te$_3$. 
\end{abstract}

\maketitle
\date{\today}

\section{Introduction}
\label{sec:intro}

The field of topological matter, both as a fundamental concept and as a search  
for materials for practical 
applications, is a very rapidly developing field~\cite{Hasan2010,Moore2010,Qi2011}. Being based on the 
physics of spin-orbit coupling (SOC), topological insulators (TIs) show 
electronic properties fundamentally distinct from conventional insulators. Their surfaces are characterized by the existence of 
metallic surface states which are protected by time-reversal and spatial 
symmetries~\cite{Fu2007,Fu2011,Hughes2011,Schindler2018}. Moreover, electrons in such states are spin 
polarized, and their spin angular moment orientation and propagation momentum are locked to each other.
This property opens promising possibilities for the generation and control of dissipationless spin currents 
that might be exploited in practical applications in spintronics, quantum computation, and thermoelectronics.

Bi-Te alloys present a wide group of materials with intriguing and technologically important properties. 
These compounds have been known and intensively studied for a long time due to their thermoelectric 
properties. In addition, recently, a growing interest in these alloys has arisen after the discovery of 
non-trivial topological insulating properties that make them very promising candidates for future generation 
of electronic devices for spintronics and quantum computation~\cite{Hasan2010}. 

Bi$_2$Te$_3$ is the first theoretically predicted and experimentally 
confirmed~\cite{Zhang2009,Chen2009,Hasan2010} prototype TI with surface states forming a single 
non-degenerate Dirac cone. It has been shown to be a strong TI (STI) 
with $\mathbb{Z}_2$ invariant (1;000) whose surface states are protected by 
time-reversal symmetry on all surfaces. It has been further shown that Bi$_2$Te$_3$ belongs to the 
family of topological crystalline insulators (TCI) 
with $n_M = -1$, where the presence of mirror symmetry leads to 
protection of the metallic surface states lying in the planes perpendicular to the mirror planes. Thus, 
Bi$_2$Te$_3$ was characterized as the first material predicted to be both a STI and a TCI. Since it exhibits 
this double topological nature, it was termed a dual TI~\cite{Rauch2014}. With such a combination, one 
can potentially exploit the fact that controlled symmetry breaking would destroy certain surface states 
while keeping others intact.

In the quest for materials with targeted topological properties, an interesting direction of investigation 
is to combine layers of known topological compounds to form heterostructures~\cite{superlatticeBiSe}. For instance, an interface 
built by a bilayer (BL) of Bi(111) on a Bi$_2$Te$_3$ substrate has been studied as a prototype system that 
indeed combines properties of the two-dimensional
(2D) and three-dimensional (3D) TIs such that topologically protected 1D edge and 2D surface 
states coexist at the surface~\cite{Hirahara2011}.

In this context, it becomes interesting to study compounds that would combine the Bi bilayers (Bi$_2$) and 
Bi$_2$Te$_3$ quintuple layers to create new stoichiometric bulk Bi-Te alloys. Bi$_1$Te$_1$, one such 
combination, was recently synthesized, characterized, and shown to be a dual 3D TI in which a weak TI 
phase and TCI phase appear simultaneously~\cite{Eschbach2017}.

Bi$_4$Te$_3$ is another well-known natural superlattice of Bi$_2$ and Bi$_2$Te$_3$ 
(see Fig.~\ref{fig:crystal}) whose topological properties are not yet thoroughly understood. In contrast to 
Bi$_2$Te$_3$ and Bi$_1$Te$_1$, this compound is considered to be a topological semimetal according to the 
Topological Materials Database~\cite{catalogue}. Bi$_4$Te$_3$ has been seen to undergo a number of 
pressure-induced structural transformations to metallic phases that lead to distinct superconducting 
states~\cite{Jeffries2011}. Saito \textit{et al}.~\cite{Saito2017} describe the Bi$_4$Te$_3$ bulk crystal taking SOC into account as 
a zero bandgap semimetal with a Dirac cone at the $\Gamma$ point. 
In all mentioned works, the theoretical analysis of the band structure was performed using density 
functional theory (DFT). By contrast, we go beyond DFT in this study and provide an analysis of the bulk and 
surface electronic structure of Bi$_4$Te$_3$ by employing the state-of-the-art quasiparticle $GW$ approach 
based on many-body perturbation theory (MBPT). This technique was applied earlier to study the 
electronic properties of other topological materials, and the importance of quasiparticle effects has been established for these 
materials~\cite{kioupakis2010,yazyev2012,nechaev2012,aguilera2013-1,nechaev2013,
Aguilera2013,rusinov2013-2,michiardi2014,nechaev2015,aguilera2019}. The results for the constituents of 
Bi$_4$Te$_3$ (i.e., Bi$_2$Te$_3$~\cite{michiardi2014} and Bi~\cite{aguilera2015}) also proved that the 
$GW$ method significantly improves the results provided by DFT. In particular, in Ref.~\cite{aguilera2015}, 
we showed with the example of Bi, the difficulties in accurately describing the electronic structure of 
semimetals or very narrow bandgap semiconductors, and we highlighted the importance of using methods beyond 
DFT to study these kinds of materials accurately. Since Bi$_4$Te$_3$ contains layers of Bi, it is not 
surprising that one needs to resort to the $GW$ method to predict the fine details of its electronic 
structure. In our approach, we perform $GW$ calculations for the bulk, and in order to compute the 
surface electronic band structure, we further parameterize a tight-binding (TB) Hamiltonian with the help 
of Wannier functions~\cite{wannier90}. The TB Hamiltonian is obtained by following the method described in 
Ref.~\cite{aguilera2019}, and thus it is calculated fully ab initio without the need for adjustable 
parameters. The calculation of bulk and surface states as well as surface resonances with this method allows us to 
provide a direct comparison between the computed spectra and results from the angle-resolved 
photoemission (ARPES) studies.

The structure of this work is as follows. We start with the description of the crystal structure of 
Bi$_4$Te$_3$ in Section~\ref{sec:crystal}. Then we provide details of the theoretical (Sec.~\ref{sec:comp}) 
and experimental (Sec.~\ref{sec:exp_details}) setups. In Section~\ref{sec:bands_bulk}, we characterize the 
bulk electronic structure and compare it with previous works. Then, in Section~\ref{sec:bands_surf}, we 
present the surface band structure and analyze the dual topological nature of Bi$_4$Te$_3$, including its 
$\mathbb{Z}_2$ topological index. In Section~\ref{sec:arpes}, we perform a comparative analysis of the 
theoretical and experimental results. Finally, we conclude in Section~\ref{sec:summary}.

\section{Crystal structure}
\label{sec:crystal}

\begin{figure}
    \centering
    \includegraphics[width=0.9\columnwidth]{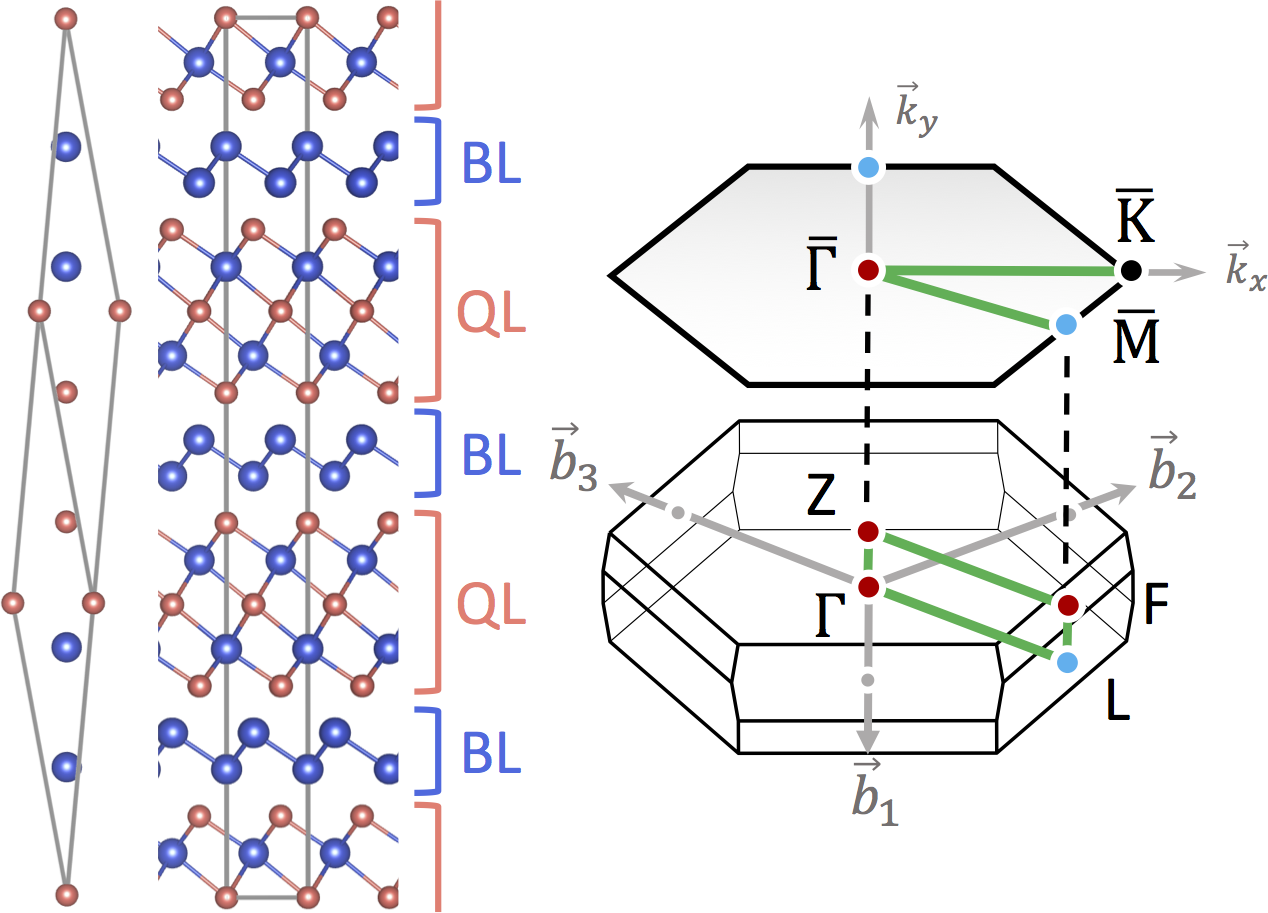}
    \caption{Crystal structure of Bi$_4$Te$_3$, consisting of 3 quintuple Bi$_2$Te$_3$ layers intercalated 
    with 3 Bi$_2$ layers, presented in the hexagonal (conventional) and rhombohedral (primitive) unit 
    cells. The Brillouin zones of the bulk crystal and the (111) surface are shown in the right panel. 
    Time-reversal invariant special $\mathbf{k}$ points are shown with red (plus) and blue (minus) dots 
    reflecting the valence state parity used to compute the topological invariant $\mathbb{Z}_2$.}
    \label{fig:crystal}
\end{figure}

The lattice structure of bulk Bi$_4$Te$_{3}$ is built as a stack of quintuple layers (QLs) of Bi$_2$Te$_3$ 
intercalated with Bi bilayers with a periodic crystal structure that can be expressed as
[(Bi$_2$)(Bi$_3$Te$_3$)]$_3$. Both conventional and primitive unit cells are shown in 
Fig.~\ref{fig:crystal}. The primitive cell reflects the rhombohedral crystal symmetry (space group R-3m) 
and consists of 7 atoms (with 2 Te and 2 Bi non-equivalent atoms).
We employ the experimental lattice parameters of~\cite{Yamana1979}. The primitive cell is determined by 
$a_{\mathrm{rho}}=14.197$\,\AA{} and $\alpha=18.0374^{\circ}$, or using the conversion to the conventional 
(hexagonal) cell $a_{\mathrm{hex}}=a_{\mathrm{rho}}\sqrt{2\,(1-\cos\,\alpha)}=4.451$\,\AA{} and 
$c_{\mathrm{hex}}=a_{\mathrm{rho}}\sqrt{3\,(1 + 2\,\cos\,\alpha)}=41.888$\,\AA{}.  
An optimization of the crystal structure (for details, see the Supplemental Material~\cite{suppmatBi4Te3}) 
has revealed the lattice constants to be very close to the experimental ones: 
$a^{\mathrm{opt}}_{\mathrm{rho}} = 14.091$\,\AA{} and $\alpha^{\mathrm{opt}} = 18.1385^{\circ}$. 
The optimized structure has a slightly (0.1\,\AA) smaller interlayer distance, which, keeping in mind the 
van der Waals type of bonding, has only a little influence on the electronic band structure (see Fig.~1 of 
the Supplemental Material~\cite{suppmatBi4Te3}). Hence, in the following, the results are obtained with the 
experimental lattice parameters given above.

\section{Computational details}
\label{sec:comp}

The electronic quasiparticle band structure is computed using the $G_0W_0$ approximation of the MBPT. 
The calculations are performed employing the code \textsc{spex}~\cite{Friedrich2010}, which implements the 
$GW$ technique in the full-potential linearized augmented plane-wave (FLAPW) framework. 
The starting point for the $GW$ calculations is provided through an interface to the FLAPW DFT code \textsc{fleur}~\cite{fleur}. 
The generalized gradient approximation (GGA) in the PBE parametrization~\cite{Perdew1996} is chosen for 
the exchange-correlation potential. 
The core electrons are treated fully relativistically by solving the Dirac equation with the spherically averaged effective potential around each nucleus. 
For the valence electrons, space is partitioned into muffin-tin (MT) spheres and an interstitial region; in the former we use an angular momentum cutoff of $l_{\mathrm{max}}=10$ and in the latter a plane-wave cutoff of 3.6~Bohr$^{-1}$. 
The MT radii are set to 2.8~Bohr for all atoms.
An 8$\times$8$\times$8 $\mathbf{k}$-point grid was used to sample the Brillouin zone.

The quasiparticle corrections are computed with the inclusion of SOC, which is known to have an important impact on the electronic properties of topological materials such as Bi$_2$Te$_3$~\cite{Aguilera2013}. 
SOC is already included in the groundstate Kohn-Sham (KS) calculations, the KS eigenfunctions are thus single-particle spinor wavefunctions. 
We employ the second-variation technique~\cite{Li1990}, in which the SOC Hamiltonian is setup and diagonalized in the basis of KS states that have been precomputed without SOC.
Specifically for this basis, we use the 400 KS states of lowest energy, which, in combination with other computational parameters, results in KS eigenenergies to be converged within 2~meV.

\textsc{spex} makes use of the auxiliary mixed product basis for representing matrix elements of the non-local operators. 
Based on our earlier experience~\cite{aguilera2013-1} as well as performed convergence tests, we use the following parameters for the $GW$ calculation.
The mixed product basis is constructed by setting the maximal angular momentum for the muffin-tin spheres to $L=5$ and $G_{max}=3.0$~Bohr$^{-1}$ for the interstitial part. 
For an accurate representation of the unoccupied states in the FLAPW method, additional basis function (so-called local orbitals) are included, two functions in each angular momentum channel with $l \leq 3$. 
A total number of about 1000 states (giving rise to about 880 empty states) corresponding to a maximal cutoff energy of $\approx 86$~eV are included for computing the screened Coulomb potential and the correlation self-energy. 
The Brillouin zone (BZ) integration is performed on an $8\times8\times8$ Monkhorst-Pack $\mathbf{k}$-point mesh. 
The contour deformation technique is used for the frequency convolution integration with
29 non-equidistant frequency points up to an (imaginary) energy of 272~eV along the imaginary axis. The self-energy is evaluated on a uniform grid with an increment of 0.54~eV along the real axis. Between the grid points, we employ spline interpolation. In this way, the self-energy is given as a continuous function along the real frequency axis, which allows us to solve the quasiparticle equation, which is nonlinear in frequency, without resorting to a linearization of the self-energy.
The $GW$ computational setup has been checked to provide the quasiparticle energies converged to within 15 meV. 
It should be mentioned that a 6$\times$6$\times$6 $\mathbf{k}$-point grid would be sufficient to reach reasonable energy convergence.
However, such a grid turned out to be too coarse for an accurate Wannier interpolated band structure.

The Wannier-function based TB Hamiltonian is used further as an input to compute the surface projected electronic band structure employing the \textsc{WannierTools}~\cite{WU2017} program package.
In this work, we are interested in a narrow region of bands lying close to the Fermi level. These bands have a predominant $p$ character, which motivates us to set $p$-type ``initial-guess'' Wannier functions for each inequivalent atom and perform the maximal localization procedure via an interface with the Wannier90 library~\cite{Pizzi2020}.

\section{Experimental details}
\label{sec:exp_details}

\begin{figure}
    \centering
    \includegraphics[width=1.0\columnwidth]{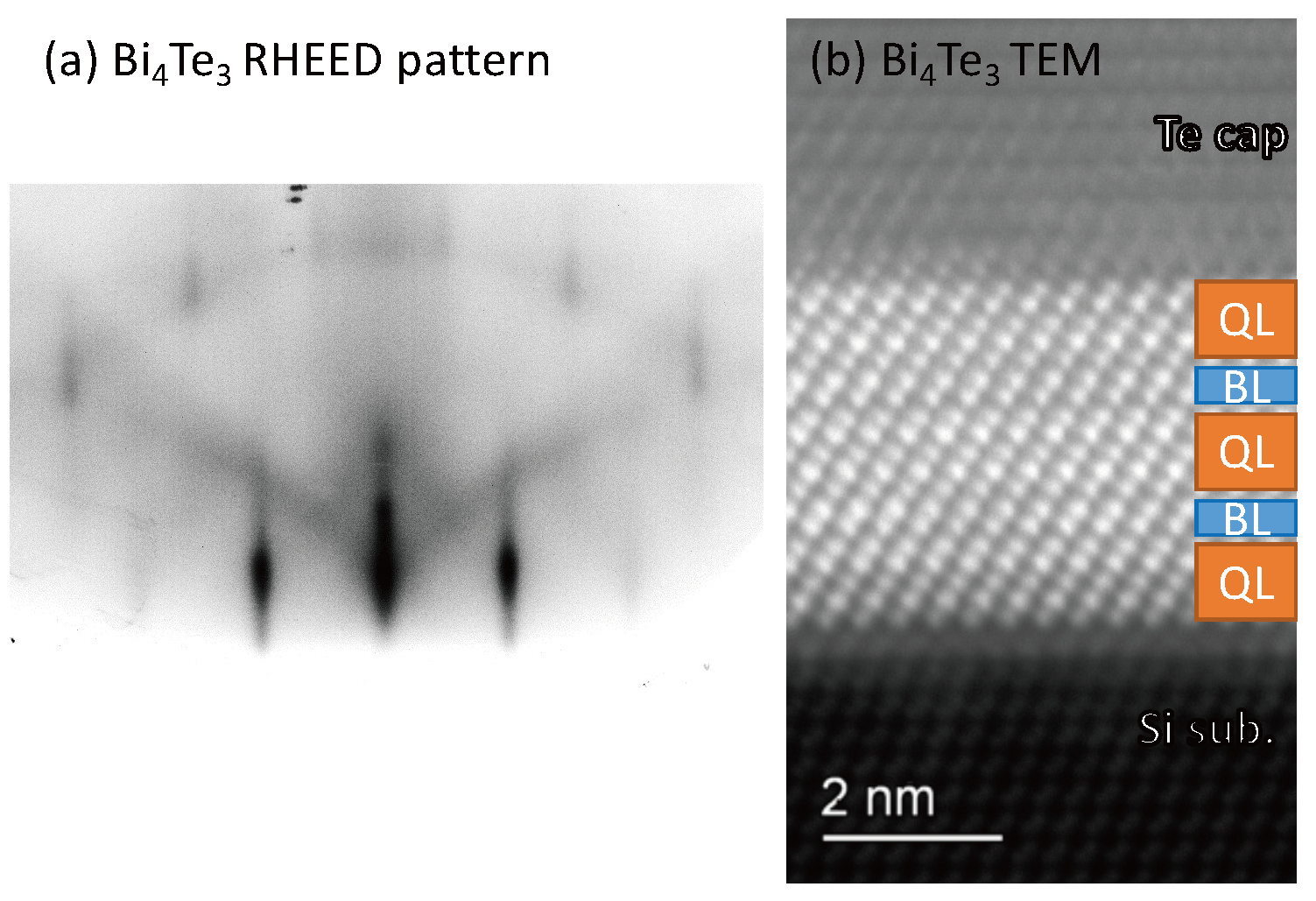}
    \caption{RHEED pattern (a) and TEM image (b) of the Bi$_4$Te$_3$(111) film with the QL termination.}
    \label{fig-sample}
\end{figure}

The samples were fabricated by the following procedure. First a clean Si(111)-$7\times7$ surface was 
prepared on an $n$-type substrate by a cycle of resistive heat treatments. Then, Bi was deposited on the 
$7\times7$ structure at 250$^\circ$C under Te-rich conditions. Such a procedure is reported to result in 
a high-quality Bi$_2$Te$_3$(111) film formation 
\cite{BiTe}. Then, the grown films were annealed at 310$^\circ$C for 
2 hours in ultra-high vacuum. During this process, Te was desorbed from the sample, and thus the sample 
became a Bi$_4$Te$_3$(111) film with the QL termination as confirmed by the transmission electron 
microscopy (TEM) observations. Figure \ref{fig-sample} shows the reflection high-energy electron diffraction (RHEED) pattern (a) and the TEM image of a typical sample. The in-plane lattice constant was determined as 4.45 $\pm$ 0.05~\AA{} and the films were determined as two-unit-cells thick with an additional QL at the interface between the film and the substrate (or the wetting layer). Details of the sample preparation and structure characterization will be 
published elsewhere~\cite{kusaka}. The ARPES measurements were performed {\it in situ} after the sample 
preparation with a commercial hemispherical photoelectron spectrometer equipped with angle and energy 
multidetections (MBS A1) at UVSOR BL-7U of UVSOR-III with $h\nu=20$~eV photons \cite{kimura}. The sample was cooled down to 16~K 
for the measurements.

\section{Results}
\label{sec:results}

\subsection{Bulk band structure}
\label{sec:bands_bulk}

\begin{figure}
    \centering
    \includegraphics[width=1.0\columnwidth]{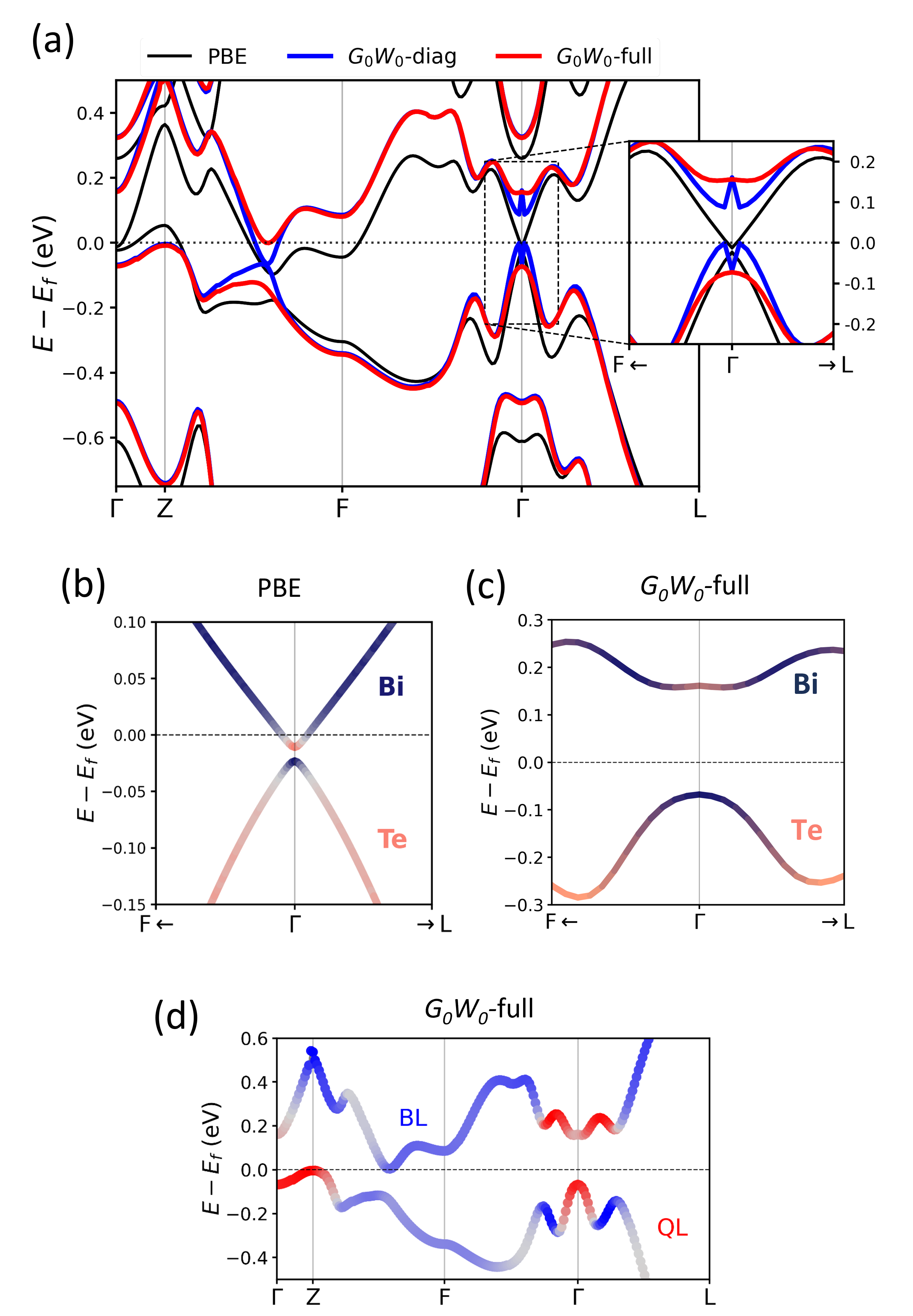}
    \caption{(a) Electronic band structure of Bi$_4$Te$_3$ in the vicinity of the Fermi energy as computed 
    with different methods: ``diag" refers to the standard $G_0W_0$ computational approach, and ``full" 
    are the results obtained by including the off-diagonal matrix elements of the self-energy. The band structures are calculated explicitly on a fine mesh along the $\mathbf{k}$ path, \emph{not} with Wannier interpolation. The bands are 
    aligned with respect to the Fermi energy. The inset shows the behavior of the valence and conduction 
    states in the vicinity of $\Gamma$. Panels 
    (b) and (c) show the band inversion between the valence and conduction states at the $\Gamma$ point calculated with PBE and $G_0W_0$-full, respectively. The color of the lines represents the contribution 
    of Bi and Te atoms. 
    (d) $G_0W_0$-full valence and conduction bands projected onto the BL (blue) and QL (red).}
    \label{fig:bands}
\end{figure}

The electronic band structure of Bi$_4$Te$_3$ is presented in Fig.~\ref{fig:bands}. The PBE band structure indicates that Bi$_4$Te$_3$ is a semimetal with a wide electron pocket along the Z--F $\mathbf{k}$-point path and a small hole pocket around the Z point. The valence (VB) and conduction 
(CB) bands are closely approaching each other at $\Gamma$, where a band inversion takes place [see Fig.~\ref{fig:bands}(b)]. In contrast to our results, the band structure presented in the 
Topological Materials Database~\cite{catalogue} seems to exhibit a crossing of the VB and CB along the $\Gamma$--Z in 
the vicinity of the Fermi level. 
On the other hand, our DFT band structure is in good agreement with the one of Ref.~\cite{Saito2017}. 
Our calculation yields an energy gap of 10~meV between the VB and CB at the $\Gamma$ point, and the CB minimum at this point is located slightly below 
the Fermi energy. The value of the gap turns out to be quite sensitive to the computational parameters. 
As it is shown in the Supplemental Material~\cite{suppmatBi4Te3}, the magnitude of the gap depends on 
the choice of the local 
orbital basis functions employed in the FLAPW method.

\begin{figure}[tbhp]
    \centering
    \includegraphics[width=0.95\columnwidth]{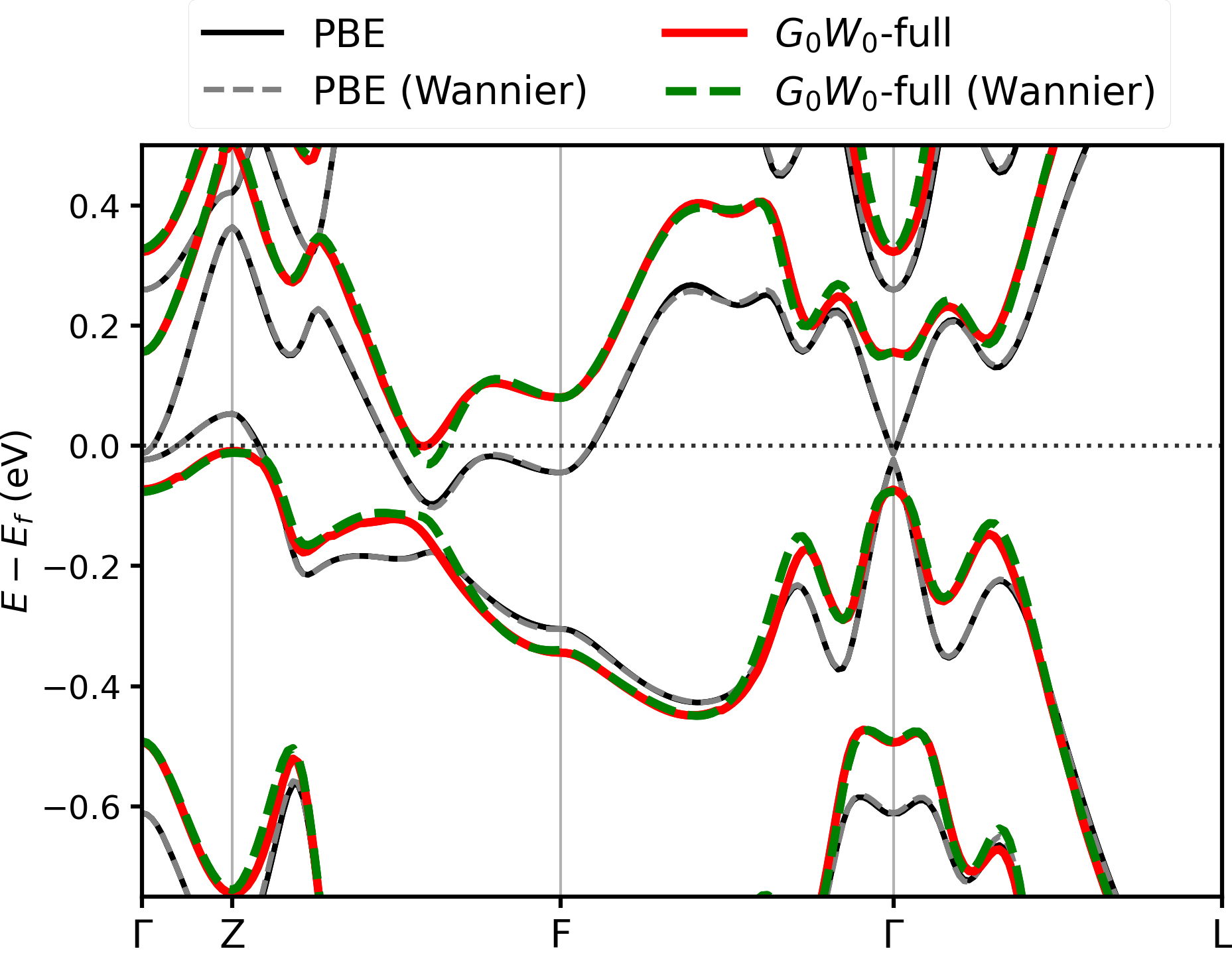}
    \caption{Comparison of the explicit and the Wannier-interpolated electronic band structures of Bi$_4$Te$_3$ in the 
    vicinity of the Fermi level.}
    \label{fig:bands-wan}
\end{figure}

Adding the quasiparticle corrections following the standard $G_0W_0$ approach by taking into account only the 
diagonal matrix elements of the self-energy (blue lines, labeled ``diag" in Fig.~\ref{fig:bands}) modifies the 
described behavior of the KS bands significantly. The quasiparticle band structure is computed on a fine $\mathbf{k}$ mesh along the 
BZ path explicitly. (For comparison, we present the band structure obtained with the Wannier 
interpolation~\cite{Pizzi2020} in Fig.~\ref{fig:bands-wan}.) 
The gap between the VB and CB at $\Gamma$ strongly increases and reaches a value of $\approx$~0.2~eV. 
Overall, the CB is shifted up in energy in the entire BZ except in a tiny area around the middle of the Z--F 
line where a narrow electron pocket remains. In contrast, the VB does not show a uniform energy shift. 
It is shifted below the Fermi level along $\Gamma$--Z. In this way, the hole pocket predicted by PBE 
disappears. 
At the same time, the VB is shifted up closer to the Fermi level around the middle of the Z--F path where it 
almost touches the CB.

Looking closer at the behavior of the VB and CB at $\Gamma$ [inset of panel (a) 
of Fig.~\ref{fig:bands}], one notices a 
characteristic anomaly: both bands show sharp narrow spikes in the energy dispersion. This unphysical 
anomaly has been observed earlier in the $G_0W_0$ band structure of the closely related topological material 
Bi$_2$Te$_3$~\cite{aguilera2013-1}. 
The anomaly is caused by the neglect of off-diagonal elements of the self-energy matrix, which play an 
important role in these materials due to strong hybridization effects caused by the SOC operator. The relativistic mass enhancement and the SOC are responsible for the band 
inversion of the VB and CB in the vicinity of $\Gamma$. The band inversion manifests itself in a mixture 
of electronic states of different orbital nature as can be seen from the atom-projected band structure shown in 
Figs.~\ref{fig:bands}(b) and (c). 
In the vicinity of the BZ center, the VB is predominantly formed by the $p$ states of Te (light pink), whereas the CB 
consists of $p$ states of Bi (dark blue). 
But in a small region around $\Gamma$, the VB and CB orbital characters are reversed. 
Due to the different band dispersions in $G_0W_0$ around $\Gamma$, the reversal of orbital character takes place in a different region of reciprocal space, much smaller in PBE [panel (b)] than in $G_0W_0$ [panel (c)]. Thus, for a proper description, the off-diagonal self-energy matrix elements need to be taken into account, because only then can the quasiparticle wavefunctions be different from the KS eigenfunctions. 
Neglecting this state mixing in $G_0W_0$ by treating only diagonal elements of the self-energy ($G_0W_0$-diag)  
leads to the unphysical behavior of the quasiparticle bands around $\Gamma$ shown as the blue lines in Fig.~\ref{fig:bands}(a). 
As is evidenced from the red lines instead (labeled ``full"), the inclusion of the 
off-diagonal matrix elements of the self-energy immediately recovers a physically meaningful dispersion 
of the bands close to $\Gamma$.
Another important modification that is induced by the treatment of the off-diagonal components is along 
the Z--F path. Instead of an effective attraction like in the case of $G_0W_0$-diag, the VB and CB 
repel each other such that the electron pocket (and thus, the semimetallic character) disappears and a fundamental indirect bandgap of 10~meV opens between the valence band maximum at Z and conduction band minimum along Z-F.
Comparison of the red and blue curves in panel (a) thus highlights the importance of taking into account 
the hybridization caused by many-body self-energy effects to describe the electronic properties of topological materials reliably. 

Since Bi$_4$Te$_3$ consists of alternating BLs and QLs, it is also interesting to look at the corresponding 
state projections. In the $G_0W_0$-full approach~\cite{aguilera2013-1}, the quasiparticle states are 
represented as linear combinations of KS states, which allows us to analyze the orbital character of 
the quasiparticle states beyond DFT [as done in Fig.~\ref{fig:bands}(c) to visualize the band inversion]. 
As one can see in Fig.~\ref{fig:bands}(d), the absolute VB maximum as well as the local maximum at $\Gamma$ 
are formed by QL states, whereas the rest of the VB is mostly formed by BL states. 
The CB in the vicinity of $\Gamma$ is also formed predominantly of QL states, but there is a noticeable 
contribution of BL states directly at $\Gamma$.

Before we proceed with the analysis of the surface band structure based on the TB Hamiltonian, we comment on the 
performance of the Wannier interpolation. For this purpose, the PBE and $G_0W_0$-full interpolated band structures are compared in 
Fig.~\ref{fig:bands-wan} with 
corresponding results obtained from explicit calculations on a fine mesh along a high-symmetry $\mathbf{k}$-path. In both 
cases, the interpolated curves are very close to the explicit ones. However, in the $G_0W_0$-full case, there 
is a small deviation along the Z--F, where the interpolated CB forms a tiny electron pocket, whereas the explicitly calculated CB does not.
We note that the Fermi energy is calculated with the 8$\times$8$\times$8 $\mathbf{k}$-point set. Any band dispersion feature that falls between the $\mathbf{k}$ points, like the electron pocket in the Wannier-interpolated band structure, is not taken into account in the determination of the Fermi energy.

\subsection{Surface band structure}
\label{sec:bands_surf}

\begin{figure}[tbhp]
    \centering
    \includegraphics[width=1.00\columnwidth]{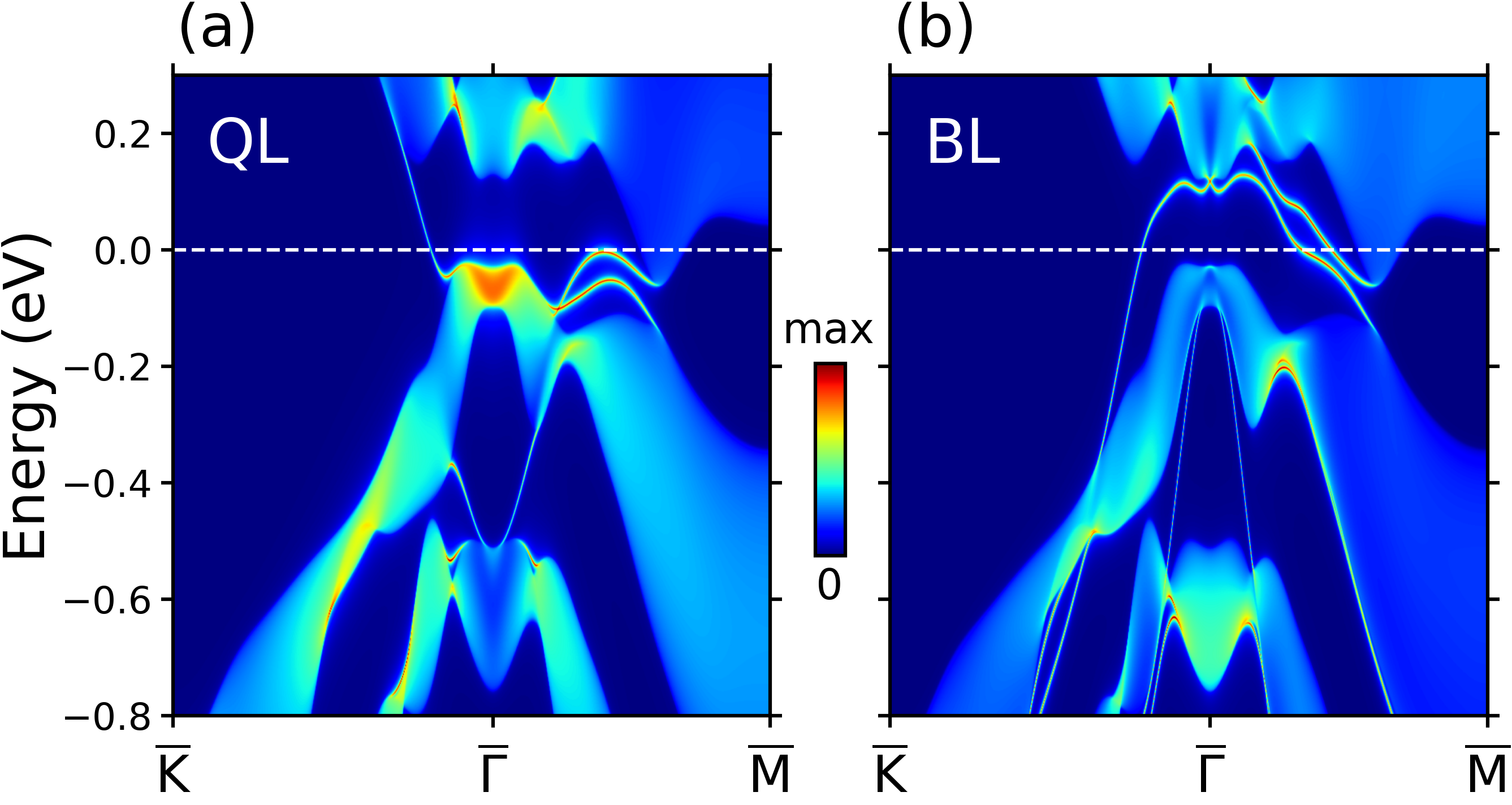}\\
    \vspace{2mm}
    \includegraphics[width=1.00\columnwidth]{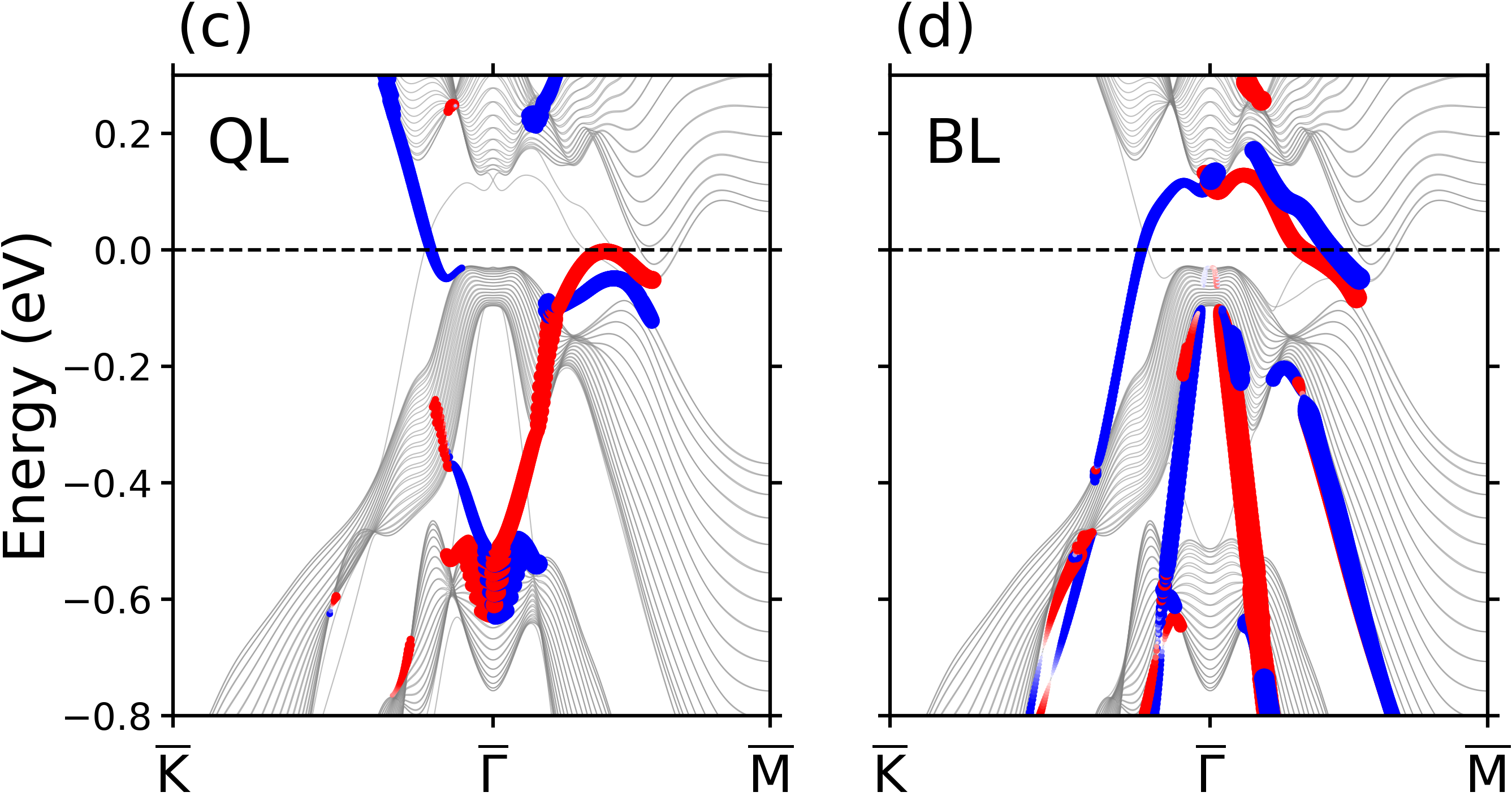}
    \caption{(a-b) $GW$-TB electronic structure of the semi-infinite Bi$_4$Te$_3$ (111) surface terminated 
    either on the QL (a) or BL (b). The color map corresponds to the density of states projected on the topmost 
    layer in each termination. Emerging surface states are represented as narrow bands arising in different 
    parts of the spectra outside of the wide bulk states. (c-d) $GW$-TB electronic structure of the 18 
    (QL+BL) layer slab. The surface states with localization either on the topmost QL (c) or BL 
    (d) levels are highlighted with colored circles. The size and color of the circles correspond to the 
    magnitude and sign (red for positive and blue for negative) of the in-plane component of the spin-polarization perpendicular to the momentum. }
    \label{fig:surface}
\end{figure}

Due to a relatively weak coupling between the QLs and BLs, the (111) plane in the rhombohedral lattice 
[(001) plane in the hexagonal definition] is the surface with the lowest surface energy. The (111) cleavage 
plane  
induces either a QL or a BL surface termination. The TB parametrization of the $G_0W_0$-full band 
structure with the help of the Wannier functions has shown to be a very powerful tool to study surface 
states of topological materials~\cite{aguilera2019}. 
Following this technique, in Fig.~\ref{fig:surface}(a-b) we present electronic band structures for a 
semi-infinite film obtained with the iterative Green's function method~\cite{lopezsancho1984} as 
implemented in {\sc WannierTools}{}~\cite{WU2017}. The color code reflects the localization rate of the 
states at the topmost surface layer such that the bright narrow bands are the emerging surface states in 
each termination. Results are shown along the surface BZ path shown in Fig.~\ref{fig:crystal}, for both 
QL and BL terminations. 
Figs.~\ref{fig:surface} (c-d) present instead the $GW$-TB electronic surface structure for a finite slab 
[18 (QL+BL) layers]. As one can see from comparison of the semi-infinite and finite spectra, the 18 
double-layer slab is sufficient to reproduce all details of the semi-infinite surface structure. This is 
in agreement with the conclusion of Ref.~\onlinecite{Chagas2020} that only a few layers are necessary to 
converge to the bulk behavior. 
The surface states with predominant localization on the lowest QL (c) and the topmost BL (d) in the slab 
are highlighted with circles whose size and color correspond to the magnitude and in-plane orientation 
(red for positive and blue for negative) of the spin-polarization perpendicular to the surface momentum. 

The surface state dispersion patterns are remarkably distinct in both terminations. 
In the QL case, the surface states are dispersing upwards in energy when going away from the BZ center, 
whereas in the BL terminations the surface states have a downward dispersion. This behavior is in 
close resemblance to the behavior of the analogous studies in Bi$_1$Te$_1$~\cite{Eschbach2017}. 
The surface states cross the Fermi level in both terminations, which already suggests a non-trivial 
topological nature of the material. Indeed, there is a single surface state crossing the Fermi level along 
the $\overline{\Gamma}$--$\overline{\mathrm{K}}$ path in both terminations. 
It is different along $\overline{\Gamma}$--$\overline{\mathrm{M}}$ where both states are located below the 
Fermi level in the QL termination, and crossing the Fermi level in the BL termination. 
It is important to point that only one surface state in the pair is connecting the VB and CB bulk bands 
and thus can be considered topologically non-trivial. 
Moreover, the topologically non-trivial band in the QL termination is approaching very closely to the Fermi 
level at its maximum along $\overline{\Gamma}$--$\overline{\mathrm{M}}$. 
Disregarding that the bulk material still has a small density of states at the Fermi energy, Bi$_4$Te$_3$ 
reveals topological features distinctive for a strong topological material. 
Since the material crystal symmetry contains an inversion center, the 
$\mathbb{Z}_2$ topological invariants are readily calculated from the valence-state parities at the time-reversal-invariant 
momenta (TRIM)~\cite{Fu2007}.
The parity invariants for each bulk TRIM ($\mathrm{K}_i$) are shown in Fig.~\ref{fig:crystal} where the 
colors of the red (blue) dots correspond to 
signs of the valence-state parity eigenvalue products 
$\delta(\mathrm{K}_a)$ = $+$ ($-1$). 
Based on the parity invariants, we obtain $\mathbb{Z}_2=(1;111)$ which characterizes Bi$_4$Te$_3$ as a 
strong topological material similar to Sb 
or Bi$_{1-x}$Sb$_x$~\cite{Teo2008}.
According to its $\mathbb{Z}_2$, one expects an odd number of surface states crossing the Fermi level and 
the presence of an electron pocket surrounding $\Gamma$, which is perfectly supported by our simulations.
Our electronic band structure differs from the one presented in the Topological Materials Database of Ref.~\onlinecite{catalogue}. 
As a consequence and in contrast to our findings, one would deduce $\mathbb{Z}_2=(0;111)$ from the data of Ref.~\onlinecite{catalogue}, which would characterize Bi$_4$Te$_3$ as a weak topological material. Therefore, experimental studies are required to back up either of the theoretical predictions.

\begin{figure}[tbhp]
    \centering
    \includegraphics[width=1.00\columnwidth]{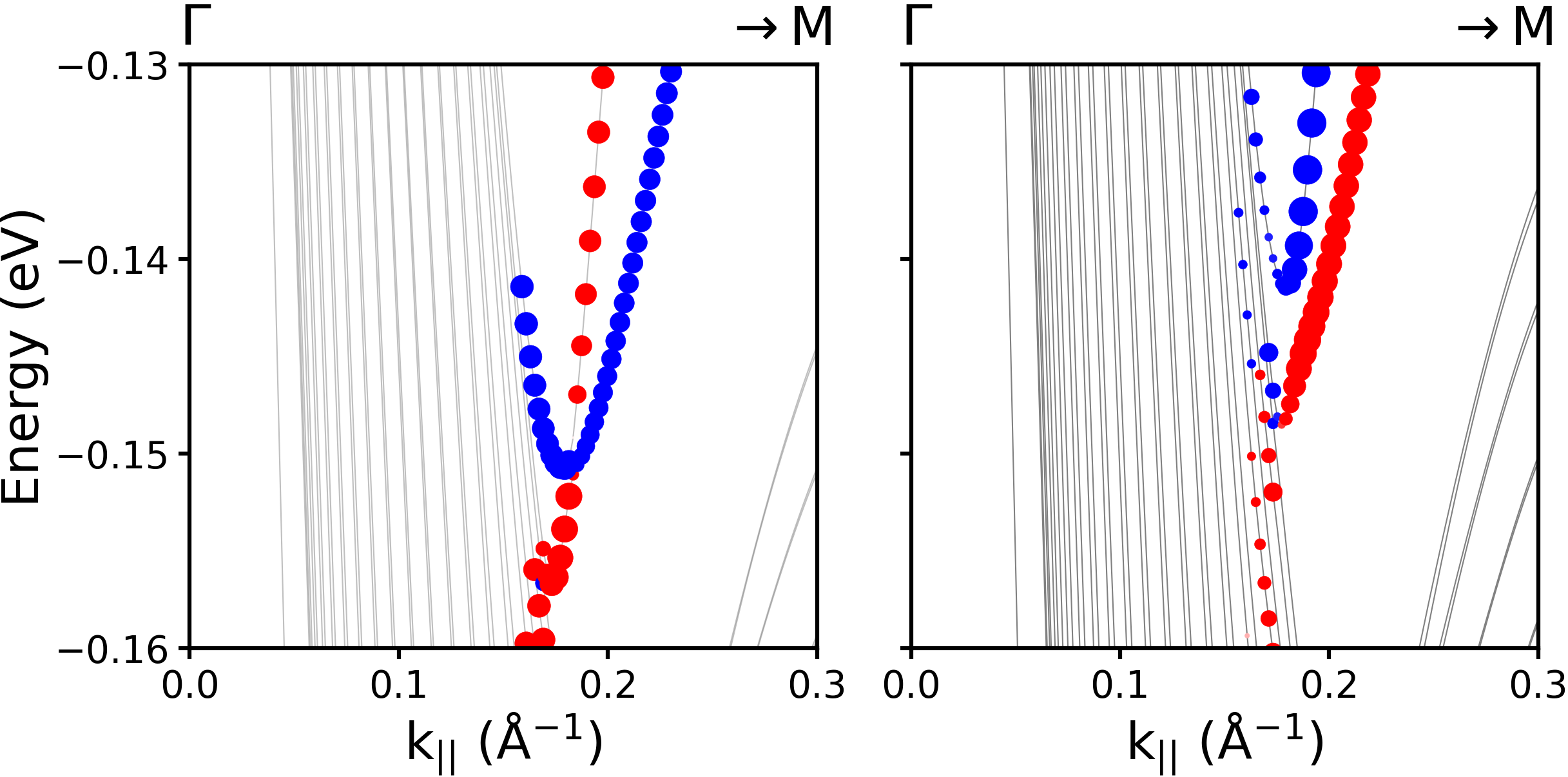}
    \caption{PBE electronic structure of the 18 (QL+BL) layer slab in the vicinity of the surface state 
    crossing point along $\overline{\Gamma}$--$\overline{\mathrm{M}}$ for the symmetric (left) and 
    deformed (right) Bi$_4$Te$_3$. Only the QL surface states are treated. The color code of the circles 
    is as in Fig.~\ref{fig:surface}.}
    \label{fig:deform}
\end{figure}

In addition to the mentioned behavior of the surface states, there are three important features that are 
revealed in Fig.~\ref{fig:surface}. First, there is a crossing between a pair of surface states in the 
QL termination along $\overline{\Gamma}$--$\overline{\mathrm{M}}$. 
The crossing is located at an energy of about $-0.1$~eV in the close vicinity of the bulk state continuum. 
This feature has motivated us to study another topological invariant that characterizes the behavior of the surface states in the presence of mirror symmetry, the magnetic mirror Chern  number $n_M$~\cite{Teo2008,Fu2011,Rauch2014}.
The result  $n_M = 1$ confirms the existence of a single surface state which connects the valence and conduction bands and can potentially lead to a crossing of the two surface bands. 
The physical picture is similar to the analogous situation in Bi$_{1-x}$Sb$_x$~\cite{Teo2008}, where the 
same value of $n_M = 1$ has been shown to be a marker of the band crossing. 
The mirror Chern number can have a non-zero value only if the crystal 
possesses a mirror plane. 
Therefore, as a check, we have slightly strained the crystal along the $\mathbf{b}_{hex}$ hexagonal axis 
and thus destroyed the mirror plane. 
In Fig.~\ref{fig:deform} we compare the behavior of the surface states in the vicinity of the discussed 
crossing point in the symmetric and deformed Bi$_4$Te$_3$ crystals. 
For demonstration purposes, we have computed the PBE surface band structures for 18 (QL+BL) layers slab. 
As it is seen from the left panel of Fig.~\ref{fig:deform}, the crossing takes place at 
$\mathrm{k}_{||}\approx0.18$~\AA$^{-1}$ in the crystal with the mirror plane. 
In the deformed crystal (right panel), the states are shifted with respect to each other such that no 
crossing occurs. 
An equivalent crossing has been observed in the surface states of Bi$_1$Te$_1$ along the same path and 
approximately at the same momentum~\cite{Eschbach2017}. 

The second feature to be discussed is a Dirac-cone like crossing located in the BL termination at $\Gamma$ 
at an energy of about 0.1~eV above the Fermi energy, slightly below the bulk conduction band minimum. 
The existence of this Dirac point for the BL termination can also be understood by looking at the system of one bilayer of Bi
deposited on Bi$_2$Te$_3$~\cite{Hirahara2011}. 
As in Bi/Bi$_2$Te$_3$, there seems to be a charge transfer from the BL to the QL. This shifts the Dirac cone above the Fermi level. There is one important consequence now in terms of the band crossing along $\overline{\Gamma}$-$\overline{\mathrm{M}}$: Since the mirror Chern number is positive, the lowest of the two spin-split surface states connects the valence band to the conduction band \cite{Teo2008}. 
When the Dirac point is below the Fermi level [QL case in Fig.~\ref{fig:surface}(c)], there is a band crossing of the two surface states. When instead the Dirac cone is above the Fermi level [BL case in panel (d)], no crossing appears.

A third Dirac-cone feature comes from the Dirac cone of Bi$_2$Te$_3$. This can be seen as a Dirac cone 
crossing at $\overline{\Gamma}$ below the Fermi level ($\approx-0.55$~eV) for the QL termination. 
The Dirac cone seems to be resonant with the bulk continuum [Fig.~\ref{fig:surface}(a)], and reveals 
itself better in Fig.~\ref{fig:surface}(c) for the finite slab as a strongly spin-polarized feature.

\section{Surface electronic structure by ARPES}
\label{sec:arpes}

\begin{figure*}[tbhp]
    \centering
    \includegraphics[width=0.95\textwidth]{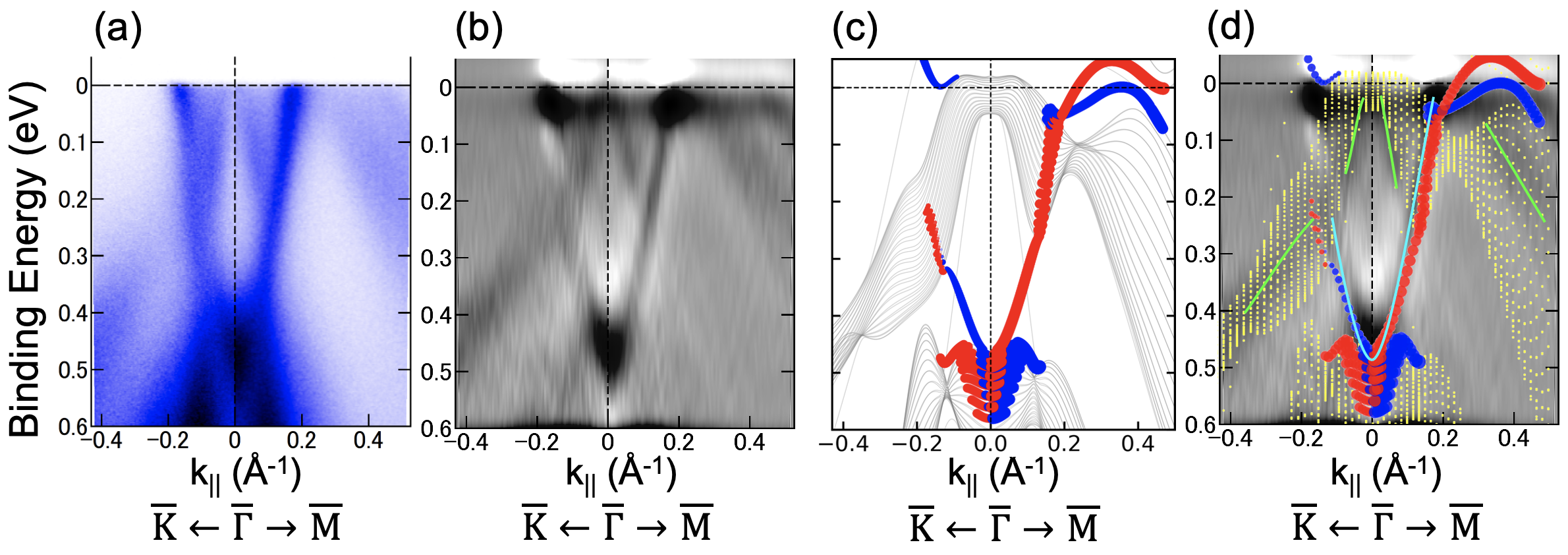}
    \caption{\label{fig:exp-1} (a),(b) ARPES spectra of Bi$_4$Te$_3$ QL terminated (111) surface. (a) is the raw data and (b) is the second derivative with respect to energy to enhance spectral features.
    (c) The same plot as in Fig.~\ref{fig:surface}(c) ($GW$-TB)  shifted 50 meV up. (d) Comparison between the 
    experimental and (shifted) theoretical electronic surface structures. Yellow dots represent computed 
    bulk states. As an eye guidance, solid lines highlight dispersion of the most intensive ARPES peaks.}
\end{figure*}
\begin{figure}[tbhp]
    \centering
    \includegraphics[width=1.00\columnwidth]{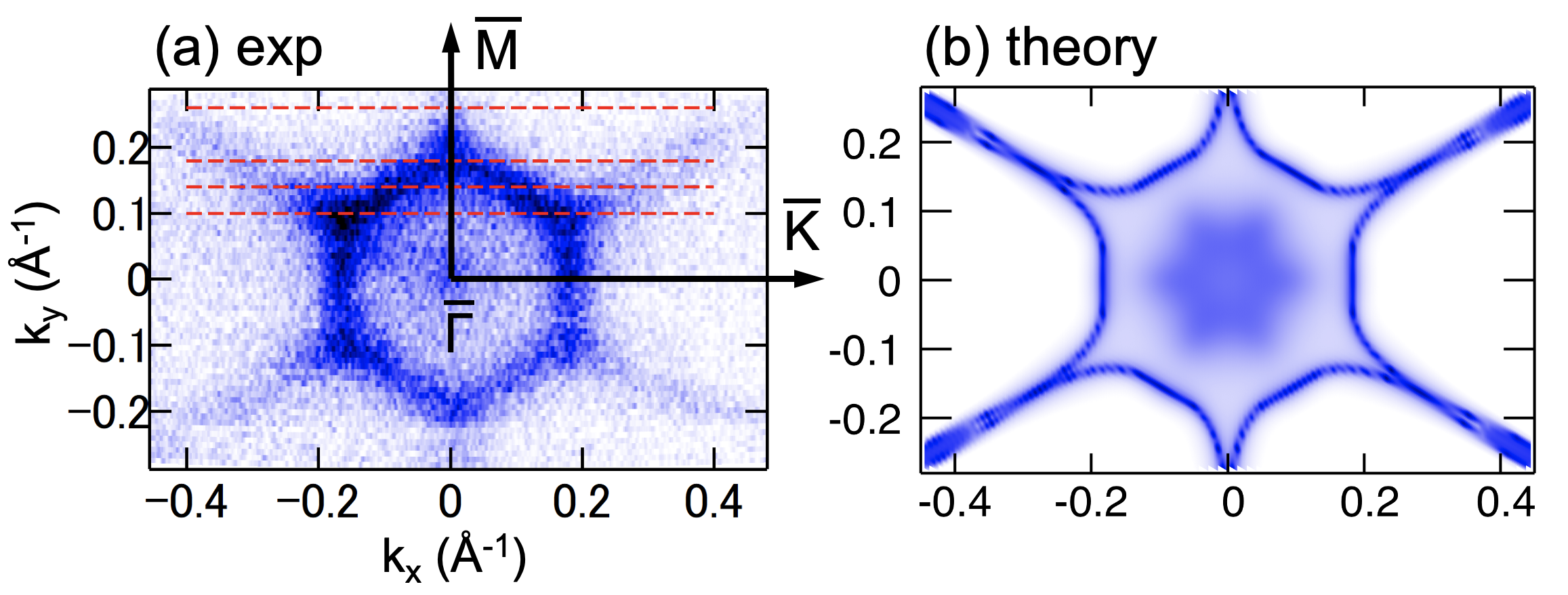}
    \caption{\label{fig:exp-2} $\mathrm{k}_{||,\mathrm{x}}$ versus $\mathrm{k}_{||,\mathrm{y}}$ (a) experimental 
    and (b) computed ($GW$-TB) Fermi surface maps. Red dashed lines show constant $\mathrm{k}_{||,\mathrm{y}}$ 
    scans presented in Fig.~\ref{fig:exp-3}.}
\end{figure}
\begin{figure*}[tbhp]
    \centering
    \includegraphics[width=1.00\textwidth]{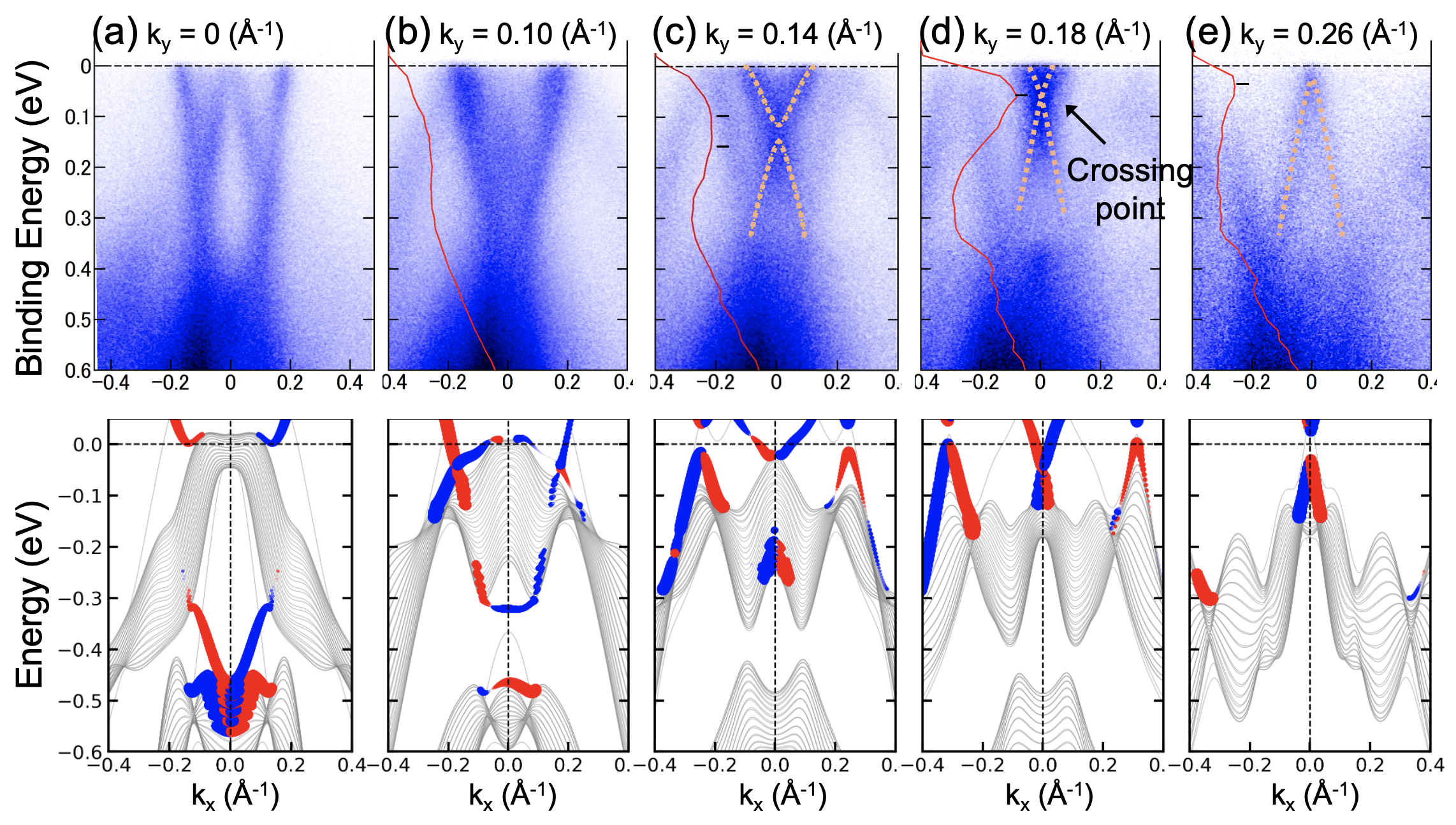}
    \caption{\label{fig:exp-3} Electronic structure of Bi$_4$Te$_3$ along the non-symmetry lines shown in Fig.~\ref{fig:exp-2}. Experimental (50 meV shifted theoretical) spectra for (a) $\mathrm{k}_{\mathrm{y}}=0$ \AA{}$^{-1}$, (b) $\mathrm{k}_{\mathrm{y}}=0.1$ \AA{}$^{-1}$, (c) $\mathrm{k}_{\mathrm{y}}=0.14$ \AA{}$^{-1}$, (d) $\mathrm{k}_{\mathrm{y}}=0.18$ \AA{}$^{-1}$, and (e) $\mathrm{k}_{\mathrm{y}}=0.26$ \AA{}$^{-1}$ are shown in the top (bottom) row. The color coding in the theoretical $GW$-TB spectra is the same as 
    in Fig.~\ref{fig:surface}(c). Dashed lines indicate the experimental surface band dispersion in the vicinity 
    of the crossing point.}
\end{figure*}

Figures~\ref{fig:exp-1}(a) and (b) present the experimental spectrum of the QL terminated (111) surface of 
a thin-film of Bi$_4$Te$_3$ as measured by ARPES for two directions. 
For comparison with the experiment, theoretical bands have been shifted 50~meV upwards in 
Fig.~\ref{fig:exp-1}(c). An overlay of the two electronic structures in Fig.~\ref{fig:exp-1}(d) shows 
the very good agreement between the experimental and theoretical spectra. 
The comparison makes it possible to distinguish the features from the bulk-like states (yellow dots) and 
from the surface states (red and blue circles). 
Thus, the experimental bands marked with green lines in Fig.~\ref{fig:exp-1}(d) correspond to bulk states. 
The sharp parabolic shape line (turquoise) is the surface state equivalent to the one that can be observed 
in Bi$_2$Te$_3$~\cite{Hirahara2011} and Bi$_1$Te$_1$~\cite{Eschbach2017}.
Note that although this band disperses sharply all the way up to the Fermi level for the $\overline{\Gamma}$--$\overline{\mathrm{M}}$ direction, its intensity is weak along the $\overline{\Gamma}$--$\overline{\mathrm{K}}$ direction at 0.1-0.2~eV due to the strong hybridization with the bulk states.
The experimental spectrum reveals also an additional band along $\overline{\Gamma}$--$\overline{\mathrm{M}}$ 
with a binding energy of $\approx30$~meV. 
As indicated in the theoretical spectrum, this is the second surface state that crosses the parabolic-like 
surface state as it was discussed in Sec.~\ref{sec:bands_surf}. 
This crossing is an important marker of the  
topological crystalline nature of Bi$_4$Te$_3$. 
To confirm that the crossing does take place, the ARPES spectra have been recorded along the non-symmetry 
lines along $\mathrm{k}_{\mathrm{x}}$ (parallel to $\overline{\Gamma}$--$\overline{\mathrm{K}}$) for different 
values of $\mathrm{k}_{\mathrm{y}}$ as is sketched in Fig.~\ref{fig:exp-2} with red dashed lines. 
These spectra are shown and compared to the corresponding theoretical band structures in Fig.~\ref{fig:exp-3}. 
As it follows from Figs.~\ref{fig:exp-3}(c),(d), and (e), the Dirac cone crossing likely occurs around 
$\mathrm{k}_{\mathrm{y}} = 0.18$~\AA{}$^{-1}$.

The Fermi surface of the thin-film Bi$_4$Te$_3$ shown in Fig.~\ref{fig:exp-2} has a hexagonal shape 
surrounding the $\Gamma$ point with straight branches going outwards at each corner. 
A similar shape is obtained in the $GW$-TB simulations. Sharp lines correspond to the crossing of the 
iso-energy surface with the surface states.
The form of the Fermi surface is as expected from the $\mathbb{Z}_2$ classification [see, e.g., Fig.~6(a) 
of Ref.~\onlinecite{Teo2008}].

\section{Summary}
\label{sec:summary}

We have performed a combined theoretical and experimental study of the electronic structure of Bi$_4$Te$_3$. 
As its closely related compounds Bi$_2$Te$_3$ and Bi$_1$Te$_1$, Bi$_4$Te$_3$ is a dual topological material: 
a strong topological insulator with $\mathbb{Z}_2$ invariant equal to (1;111), and a topological crystalline 
insulator with mirror Chern number $n_M = 1$. 
The semimetallic groundstate predicted in DFT changes into a small indirect band gap semiconductor upon 
inclusion of the quasiparticle corrections in the framework of the $G_0W_0$ approximation of many-body 
perturbation theory.

We discussed how inclusion of off-diagonal matrix elements of the electronic self-energy is important for 
capturing the fine details of the hybridization of states due to the strong spin-orbit coupling. This 
hybridization between many-body states is responsible for the gap opening in Bi$_4$Te$_3$.

Theoretical analysis of Bi$_4$Te$_3$ films reveals topologically non-trivial surface states. 
The presence and dispersion of these states are consistent with ARPES measurements of a QL terminated 
Bi$_4$Te$_3$ thin film.
Both theory and experiment evidence a Dirac crossing between the surface states in the valence region. 
This crossing is attributed to the presence of a mirror symmetry and corresponding non-trivial value of 
the mirror Chern number.

\begin{acknowledgments}
We gratefully acknowledge the computing time granted through JARA-HPC on the supercomputer JURECA at 
Forschungszentrum J\"ulich. M.T. acknowledges the kind hospitality of the Peter Gr\"unberg Institut and 
Institute for Advanced Simulation, Forschungszentrum J\"ulich at the beginning of this project.
Taisuke Sasaki and Kazuhiro Hono are acknowledged for their help in the TEM observation, and Kazuki Sumida, Kiyohisa Tanaka and Shin-ichiro Ideta are acknowledged for their help in the ARPES measurements.
This work was supported by Grants-In-Aid from the JSPS KAKENHI (Grant No. 18H03877), the Murata Science Foundation (Grant No. H30084), the Asahi Glass Foundation, the Iketani Science and Technology Foundation (Grant No. 0321083-A), and a Tokyo Tech Challenging Research Award.
The ARPES measurements were performed under UVSOR Proposals No. 19-569, and No. 19-858, and No. 20-777.
D.N. and S.B. are supported by the European Centre of Excellence MaX ``Materials design at the Exascale'' (Grant No. 824143) funded by the EU. S.B.\ is grateful for financial support from the Deutsche Forschungsgemeinschaft (DFG) through  the Collaborative Research Center SFB 1238 (Project C01).
\end{acknowledgments}

\end{document}


\title{Supplementary material for "Bulk and surface electronic structure of Bi$_4$Te$_3$ from $GW$ calculations and photoemission experiments"
}

\author{Dmitrii Nabok}
\affiliation{Peter Gr\"{u}nberg Institut and Institut for Advanced Simulation, Forschungszentrum J\"{u}lich and JARA,
52425 J\"{u}lich, Germany}

\author{Murat Tas}
\affiliation{Department of Physics, Gebze Technical University, Kocaeli 41400, Turkey}

\author{Shotaro Kusaka}
\affiliation{Department of Physics, Tokyo Institute of Technology, 2-12-1 Ookayama, Meguro-ku, Tokyo
152-8551, Japan}

\author{Engin Durgun}
\affiliation{UNAM - National Nanotechnology Research Center and Institute of Materials Science and
Nanotechnology, Bilkent University, Ankara 06800, Turkey}

\author{Christoph Friedrich}
\author{Gustav Bihlmayer}
\author{Stefan Bl\"{u}gel}
\affiliation{Peter Gr\"{u}nberg Institut and Institut for Advanced Simulation, Forschungszentrum J\"{u}lich and JARA,
52425 J\"{u}lich, Germany}

\author{Toru Hirahara}
\affiliation{Department of Physics, Tokyo Institute of Technology, 2-12-1 Ookayama, Meguro-ku, Tokyo
152-8551, Japan}

\author{Irene Aguilera}
\email[Corresponding author: ]{i.g.aguilerabonet@uva.nl}
\affiliation{Peter Gr\"{u}nberg Institut and Institut for Advanced Simulation, Forschungszentrum J\"{u}lich and JARA,
52425 J\"{u}lich, Germany}

\maketitle
\date{\today}

\section{Computational details}

\subsection{Volume optimization}

In this study, we have employed the experimental Bi$_4$Te$_3$ crystal 
structure~\cite{Yamana1979}. However, as a crosscheck, we have carried out a full crystal structure relaxation.

The structural optimizations are performed by first-principles calculations based on density functional theory 
(DFT)~\cite{kohn1965self,hohenberg1964inhomogeneous} as implemented in the Vienna $Ab~initio$ Simulation Package 
(VASP)~\cite{kresse1993ab,kresse1994ab,kresse1996efficiency,kresse1996efficient}. 
We employ the generalized gradient approximation in the parametrization of Ref.~\onlinecite{Perdew1996} for the exchange-correlation functional. 
The van der Waals interactions are included by considering the Grimme approach (DFT-D3)~\cite{grimme2010consistent}. 
The projector augmented-wave (PAW) method \cite{PhysRevB.50.17953} was used to 
describe the element potentials with an energy cut-off of 230 eV. 
The lattice constant and atomic relaxations were performed by using the conjugate gradient optimization, allowing 
10$^{-5}$~eV and 0.01 eV/\AA{} tolerance on energy (between two sequential steps) and force (on each atom), 
respectively. 
The numerical integrations over the Brillouin zone were calculated by using $12\times12\times12$ $\mathbf{k}$-points 
meshes~\cite{monkhorst1976special}.
The optimized lattice constants are provided in Table~\ref{tab:lattice}. 
\begin{table}[h]
    \centering
    \begin{tabular}{l|c|c|c|c}
            & a$_{\mathrm{rhom}}$ (\AA{}) & $\alpha_{\mathrm{rhom}}$~($^\circ$) & a$_{\mathrm{hex}}$ (\AA{}) & c$_{\mathrm{hex}}$ (\AA{}) \\
        \hline
        opt & 14.091 & 18.138 & 4.442 & 41.568  \\
        exp~\cite{Yamana1979} & 14.197 & 18.037 & 4.451 & 41.888  \\
    \end{tabular}
    \caption{\label{tab:lattice} The experimental and computationally optimized with DFT-D3 lattice parameters of Bi$_4$Te$_3$. 'rhom' and 'hex' subscripts denote the lattice constants of the primitive and conventional cells, correspondingly.}
    \label{tab:my_label}
\end{table}
\begin{lstlisting}[caption={VASP POSCAR input for the optimized DFT-D3 bulk structure of Bi$_4$Te$_3$.}, label={lst:poscar}]
Bi4 Te3
1.0
13.915068 -2.221169  0.000000
13.915068  2.221169  0.000000
13.560518  0.000000  3.830805
Bi Te
4 3
direct
0.714125 0.714125 0.714125 Bi
0.285875 0.285875 0.285875 Bi
0.146433 0.146433 0.146433 Bi
0.853567 0.853567 0.853567 Bi
0.000000 0.000000 0.000000 Te
0.422085 0.422085 0.422085 Te
0.577915 0.577915 0.577915 Te
\end{lstlisting}
The entire optimized with VASP bulk crystal structure of Bi$_4$Te$_3$ is listed in Listing~\ref{lst:poscar}.

\begin{figure}
    \centering
    \includegraphics[width=0.85\columnwidth]{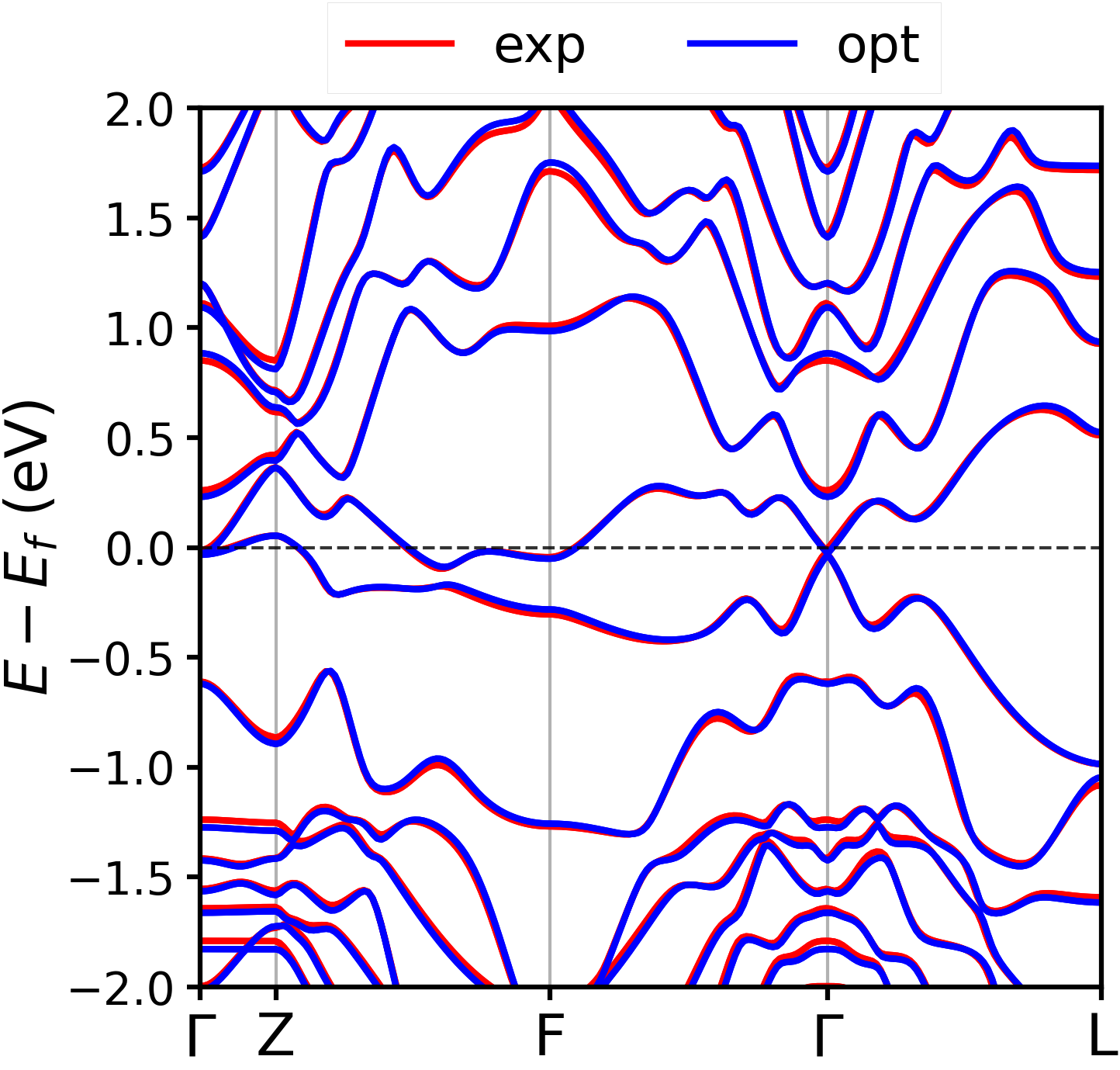}
    \caption{\label{fig:bands-exp-vs-opt} PBE+SOC band structures computed using the experimental and optimized lattices.}
\end{figure}

\begin{figure}
    \centering
    \includegraphics[width=1.0\columnwidth]{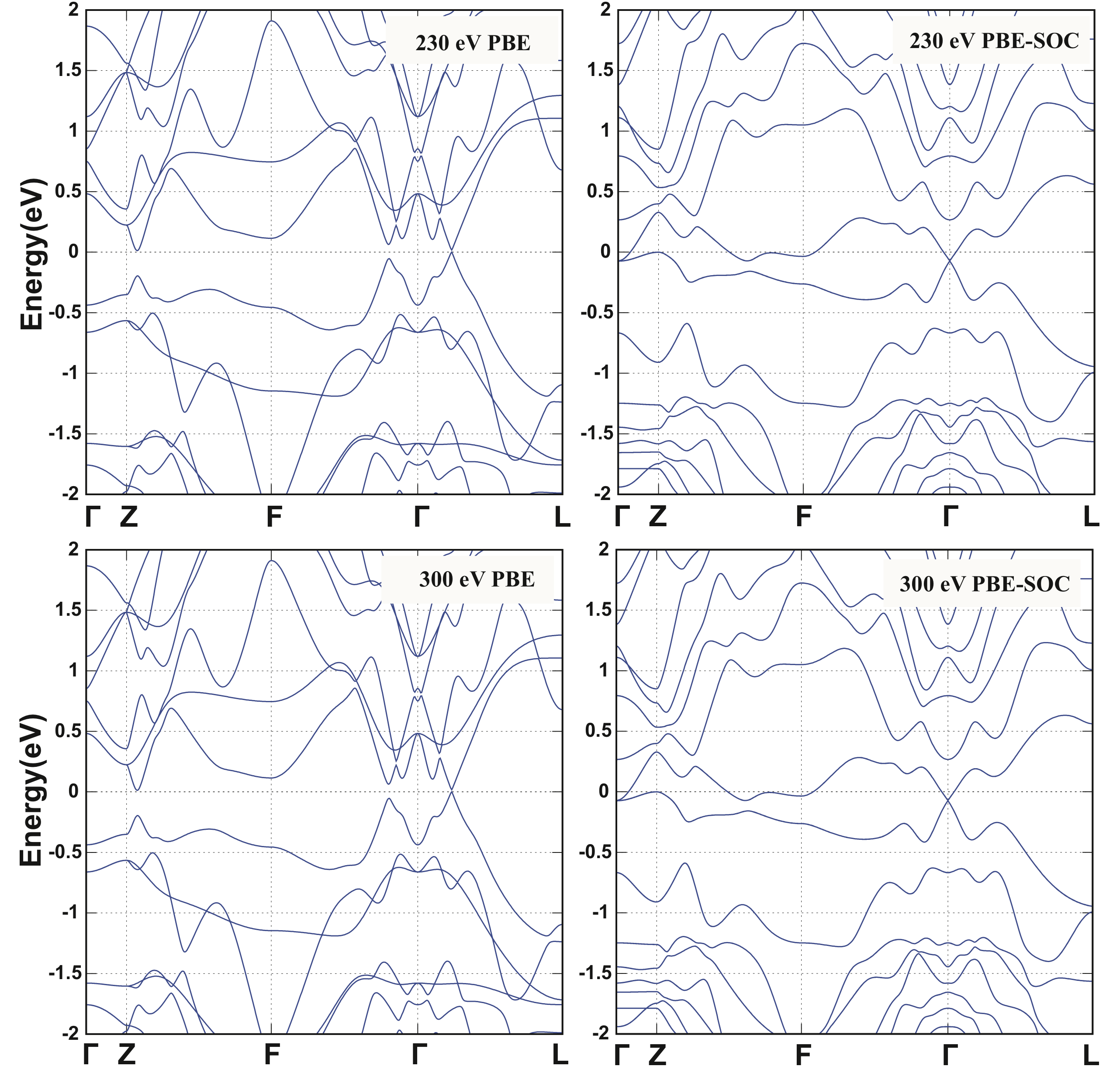}
    \caption{\label{fig:bands-vasp} Comparison of electronic structures computed with VASP for two values of the plane-wave energy cutoff (230 and 300~eV) and with and without SOC.}
\end{figure}

The PBE+SOC electronic band structures obtained for the experimental and optimized crystal volumes are shown in 
Fig.~\ref{fig:bands-exp-vs-opt}. The band structures are computed with the \textsc{fleur} code with the parameters specified in the section ``Computational details'' of the main text.
As expected, due to the very close agreement between the experimental and optimized lattice parameters, the
difference in the electronic structure is very small and does not exceed 50 meV.
For completeness, in Fig.~\ref{fig:bands-vasp}, we present the PBE and PBE+SOC band structures computed with VASP for two values of the energy cutoff 230 eV and 300 eV. The results demonstrate the convergence of the electronic band structure with respect to the basis size. The figure also highlights the importance of the SOC effects in Bi$_4$Te$_3$. Finally, the excellent agreement between the \textsc{fleur} (Fig.~\ref{fig:bands-exp-vs-opt}) and VASP (Fig.~\ref{fig:bands-vasp}) band structures supports the high accuracy of the performed calculations.

\section{Importance of the local basis functions}
\label{app:helos}

\begin{figure}
    \centering
    \includegraphics[width=1.0\columnwidth]{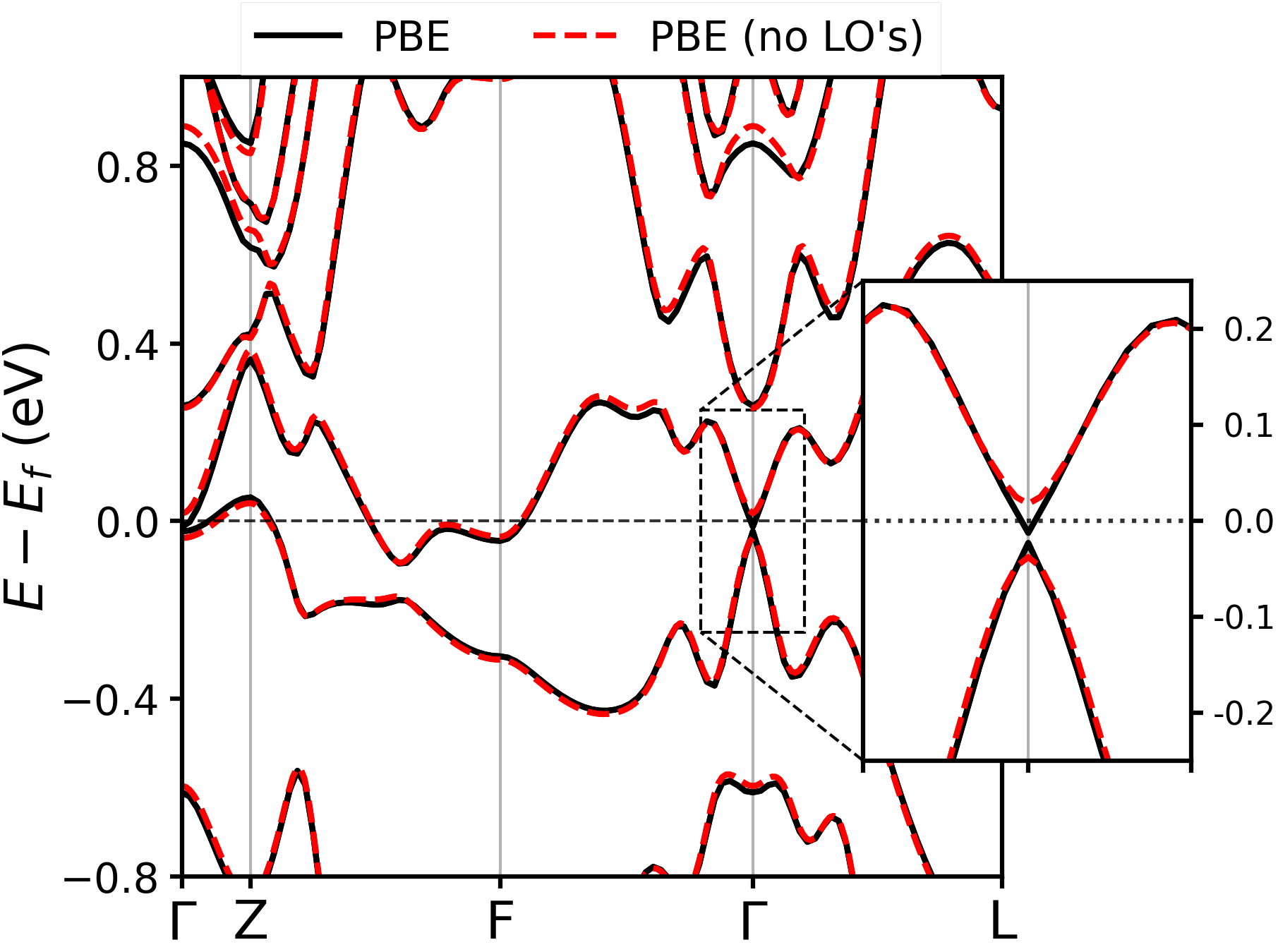}
    \caption{PBE band structures computed with and without additional local basis function.}
    \label{fig:bands-helo}
\end{figure}

As was mentioned in Section~III, the local part of the LAPW basis is complemented with extra basis functions in 
addition to those that are required by inclusion of the semi-core states of Bi and Te (and therefore included 
into the basis by default). 
The role that these extra functions play in our FLAPW calculations is twofold. 
On the one hand, the importance of those extra local orbitals is well established for $GW$ calculations in the 
framework of the FLAPW method~\cite{Friedrich2006,Nabok2016} to compute accurate quasiparticle energies. 
On the other hand, the inclusion of the additional basis function in our case turns to be already important to 
converge the Kohn-Sham electronic band structure in the vicinity of the Fermi energy.
As shown in Fig.~\ref{fig:bands-helo}, the usage of the default \textsc{fleur} basis setup 
can lead to small deviations in the band structure with respect to the fully converged case.
Tiny as they may be, they do play a role on the small energy scale of the current study. The largest deviations can be observed at the $\Gamma$ point where the effect of the spin-orbital interaction is 
most pronounced. 
Although the differences are typically rather small (as it is shown, e.g., in the inset of Fig.~\ref{fig:bands-helo}), 
they are still of relevance for the subsequent $GW$ calculations.
%

%

%